# Artificial Increasing Returns to Scale and the Problem of Sampling from Lognormals

**Andrés Gómez-Liévano[1], Vladislav Vysotsky[2] and José Lobo[3]**

**Abstract**

We show how Increasing Returns to Scale (IRS) in urban scaling can artificially emerge, systematically and predictably, without any sorting or positive externalities. We employ a model where individual productivities are independent and identically distributed (i.i.d.) lognormal random variables across all cities. We use extreme value theory (EVT) to demonstrate analytically the paradoxical emergence of IRS when the variance of log-productivity is larger than twice the log-size of the population size of the smallest city in a cross-sectional regression. Our contributions are to derive an analytic prediction for the artificial scaling exponent arising from this mechanism, and to develop a simple statistical test to try to tell whether a given estimate is real or an artifact. Our analytical results are validated analyzing simulations and real microdata of wages across municipalities in Colombia. We show how an artificial scaling exponent emerges in the Colombian data when the sizes of random samples of workers per municipality are 1% or less of their total size.

[1]Center for International Development, Harvard University, Cambridge, USA
[2]Department of Mathematics, University of Sussex, Brighton, UK
[3]School of Sustainability, Arizona State University, Tempe, USA

**Corresponding author:**
Andrés Gómez-Liévano, Growth Lab at the Center for International Development, John F. Kennedy School of Government at Harvard University, 79 JFK Street, Cambridge, MA 02138, USA.
Email: Andres Gomez@hks.harvard.edu

## Introduction

The origin of the firm-size and urban-size wage premia is still a topic of active research. The literature in both subjects shows consensus that individual wages, on average, are higher in larger firms (Oi and Idson 1999) and in larger cities (Rosenthal and Strange 2004). The question of why population size is associated with these economic advantages is still hotly debated in the urban economics literature (e.g., Duranton and Puga 2004; Hollister 2004; Lehmer and Möller 2010; Combes and Gobillon 2015; Eeckhout and Kircher 2018), and has become the foundation to think about a unified theory of urban phenomena (Pumain et al. 2006; Bettencourt and West 2010; Bettencourt 2013; Batty 2013; Martinez 2016; Gomez-Lievano et al. 2016). Our contribution to this literature is to study unexpected analytical consequences and empirical challenges posed by the simple fact that wages are approximately lognormally distributed (see Kleiber and Kotz 2003 for a review; for older discussions, see Roy 1950; Shockley 1957; Aitchison and Brown 1957; Mincer 1970). In contrast to normal distributions, lognormals can generate extremely large values. Such feature of lognormals, we show, can result in misestimating a statistical effect that can be mistaken for superlinear urban scaling.

We do not seek to re-evaluate accepted explanations for the urban and firm size productivity premia, or refute urban scaling theories. Our results are about a potential source of bias in the estimate of scaling exponents for variables with extremely large variance, which slows down convergence of sample means to population means. Our concern here is not with a bias in the sampling process, e.g. how statistical offices, researchers, or measurements devices sample information from cities. Our attention is on the estimate of the scaling exponent when studying variables like wages, even when wages are "uniformly sampled" in each city. The main consequence of our work is methodological, and our results are relevant to studies that investigate the effects of size on wages, when the latter have been aggregated into averages, or into total wage bills. Thus, our research is relevant for studies like those of Rice et al. (2006); Andersson et al. (2016); Strumsky et al. (2019); Keuschnigg et al. (2019); Keuschnigg (2019).[1]

The general issue with wages, which drives the results we present in this paper, is that they are heavy-tailed distributed and, because of that, contain outliers. Researchers have devised rules-of-thumb to deal with outliers, but a rigorous analysis has been missing. We address this presenting analysis of how outliers can influence the estimation of increasing returns to scale (IRS). We list our contributions as follows:

- We identify a *mechanism* that can generate a spurious rise in the average wage with sample size, which can be mistaken as evidence of urban scaling.
- We identify *conditions* when this mechanism is likely to occur in wage data.
- We derive an *analytic expression* for the (spurious) scaling exponent of

[1] Glaeser and Maré (2001), and in particular Combes et al. (2008), initiated the study of the urban wage premium using individual-level data. Individual data are used because they can address endogeneity concerns related to disentangling agglomeration externalities from sorting of skilled workers into large cities (see Combes and Gobillon 2015). However, the methodological switch from aggregate to individual level data leaves unspecified other potential statistical problems with aggregating variables like wages.

wages to size generated.

- We develop a simple *statistical test* to try to rule out this mechanism as an explanation of wage size premia.

Our present work can generalize to other measures of individual output for which city averages are often constructed, like average income or average number of patents. Our work, however, must be distinguished from scaling analysis that focuses on counts like homicides, cases of infectious disease, or employment, and the corresponding issues of statistical underreporting that might occur. In contrast to problems of statistical underreporting, our paper is about problems of statistical convergence of sample means.

The paper is organized as follows. First, we set the stage for our study and give a brief overview of urban scaling and the mechanisms behind it. Second, we derive the main analytical results. Third, we present numerical simulations followed by a real-world application where we analyze the superlinear scaling of wages with size in Colombian municipalities. We end with a discussion and conclusion of our results.

## Framework

In cities, IRS is typically quantified by the value of the exponent of a power-law function $F$ that relates the total output $Y$ to population size $n$, by $Y = F(n)$, where $F(n) = Y_0 n^{1+\delta}$. IRS happens when $\delta > 0$ (e.g., Sveikauskas 1975; Rosenthal and Strange 2004; Bettencourt 2013). For example, Bettencourt et al. (2007) showed that $\hat{\delta} \approx 0.12$ for total wages in U.S. Metropolitan Statistical Areas.

The scaling exponents $\beta = 1 + \delta$ and $\delta$ quantify how "elastic" is total output and output per capita, respectively, in percentage terms to a percent change in population size $n$. Thus, $\beta$ is mathematically defined as $\beta = \frac{\Delta F/F}{\Delta n/n}$. In the limit it can be written as $\beta = \frac{\mathrm{d}\ln(F(n))/\mathrm{d}n}{\mathrm{d}\ln(n)/\mathrm{d}n}$, where $\mathrm{d}/\mathrm{d}n$ is the derivative with respect to $n$ and we assume that $Y = F(n)$ is a function of $n$ only. We will sometimes refer to scaling exponents as "elasticities".

The observation that larger scales are associated with higher productivity is usually explained by one, or a combination, of two general mechanisms: productive individuals sorting themselves into large cities ("assortative matching" in the firm-size premium literature), or larger cities generating more productive individuals through positive externalities coming from their interactions.[2] We will refer to these two mechanisms simply as sorting and agglomeration effects, respectively. These effects come from specific economic processes which entail either decisions by, or interactions among, economic agents, whose absence would result in the absence of IRS. The present work refutes this claim.

While sorting and agglomeration effects have been shown to cause IRS (Glaeser and Maré 2001; Rosenthal and Strange 2004; Melo et al. 2009; Bettencourt 2013; Behrens et al. 2014; Combes and Gobillon 2015), the presence of IRS does not imply sorting or agglomeration effects. Thus, we will show that urban scaling can also emerge as the consequence of extreme values of productivity contributing

[2] There may also be selection effects that eliminate the least productive firms

significantly to the total output of the aggregate. To our knowledge, we are the first to demonstrate these paradoxical results analytically for the case of lognormal productivities. We also propose a method to tell apart real versus artificial scaling exponents in data. The method is based on a simple intuition: that randomization of individuals in the data across cities should eliminate the economic effects but not the artificial one. To state this differently, the artificial effect, if present, is exclusively due to randomness and it will be observed after randomization of the data.

In our analysis, we will abstract away any market, equilibrium condition, or coordination mechanism among individuals, since our main claim is that the presence of IRS is not necessarily evidence of sorting, coordination, interactions or positive externalities. In our model, productivities will be independently and identically sampled, and yet we will show that total output will display increasing returns for a wide range of scales.

## Analytic Results

Assume individuals, regardless of the city they live in, have productivities independently and identically distributed (i.i.d.), sampled from a lognormal distribution $\mathcal{LN}(x_0, \sigma^2)$, whose probability density function is

$$p_X(x;\, x_0, \sigma^2) = \frac{1}{x\sqrt{2\,\pi\sigma^2}} e^{-\frac{(\ln x - \ln x_0)^2}{2\sigma^2}}, \quad (1)$$

where $x_0$ and $\sigma$ are positive parameters such that $\ln(x_0) = E[\ln(X)]$ and $\sigma^2 = Var[\ln(X)]$. A simple computation yields the value of expected productivity $\mu \equiv E[X] = x_0 e^{\sigma^2/2}$. We will use upper case and lower case letters to denote random variables and their possible values, respectively.

We define the total output of a city of population $n$ as the sum of the productivities of its inhabitants, $Y(n) = \sum_{i=1}^{n} X_i$. Henceforth, we will assume that there are $m$ cities, indexed as $k = 1, \dots, m$, each with total populations $n_1, \dots, n_m$.

The choice of a lognormal distribution has two purposes. First, there is evidence that the empirical distributions of productivity across workers, such as wages in cities, are well-described by lognormal distributions (Roy 1950; Aitchison and Brown 1957; Mincer 1970; Kleiber and Kotz 2003; Combes et al. 2012; Eeckhout et al. 2014). Second, despite all its moments being finite, the lognormal distribution has a property which enables the emergence of IRS as an artificial effect: namely, lognormals are heavy-tailed which tend to generate extremely large positive values due to very high variance.

### *Elasticity for a single city*

Let us proceed by calculating first the change in the expected value of urban output according to the above simple model if population size is increased by $\lambda > 1$:

$$\mathrm{E}[Y(\lambda n)] = \mathrm{E}\left[\sum_{i=1}^{\lambda n} X_i\right], \quad (2)$$

$$= \sum_{i=1}^{\lambda n} \mathrm{E}[X_i],$$
$$= \lambda n \, \mathrm{E}[X_1],$$
$$= \lambda \mathrm{E}[Y(n)].$$

Dividing both sides by $\lambda n$ we get per capita terms,

$$\mathrm{E}\left[\frac{Y(\lambda n)}{\lambda n}\right] = \mathrm{E}\left[\frac{Y(n)}{n}\right].$$

From the point of view of expectation values, our model does not display IRS, and the expected per capita output is constant across cities. Specifically, the total expected production in our model is $E[Y(n)] = Y_0 n^{\beta}$, with the scaling exponent $\beta = 1$ and $Y_0 = \mu$. While the derivation of equation (2) might seem trivial, the fact that $E[Y(n)]$ is never observable is not so obvious. What we actually observe is the realized $Y(n)$, a crucial distinction when the distribution $X_i$ has certain properties. We must go beyond relying on expectation values and study how the distribution of $X_i$ determines whether $Y(n)$ displays IRS with a superlinear exponent.

Our approach to understand how a stochastic variable like $Y(n)$ scales with sample size $n$ draws on the probabilistic theory of the so-called "stable laws". This theory gives insights about finding sequences $c_n$ and $d_n$ such that the probability distribution function (CDF) $\Pr\left(c_n^{-1}(Y(n) - d_n) \leq x\right)$ converges to that of a stable distribution. When we find such sequences, we can state that $Y(n)$ scales with $n$ as $d_n$ does. Thus, the scaling exponent $\beta$ can be computed using $d_n$, as

$$\beta(n) = \frac{\mathrm{d} \ln(d_n)/\mathrm{d}n}{\mathrm{d} \ln(n)/\mathrm{d}n}, \tag{3}$$

where we have made explicit a possible functional dependence on $n$, anticipating that the scaling exponent may be affected by sample sizes. To see why this approach is useful, let us discuss the situation when the Central Limit Theorem (CLT) holds. According to the CLT, when $X_i$ are i.i.d. with finite mean and variance, and $n$ is very large, then the stable law to which $\Pr\left(c_n^{-1}(Y(n) - d_n) \leq x\right)$ converges to as $n \to \infty$ is the Standard Gaussian distribution if $d_n = E[X_1]n$ and $c_n = (Var[X_1]n)^{1/2}$. Thus, denoting $E[X_1] = \mu$, as $n \to \infty$, the scaling exponent of total output with respect to size is

$$\begin{aligned} \beta &= \frac{\mathrm{d} \ln(d_n)\,/\mathrm{d}n}{\mathrm{d} \ln(n)\,/\mathrm{d}n} \\ &= \frac{\mathrm{d} \ln(\mu\, n)\,/\mathrm{d}n}{\mathrm{d} \ln(n)\,/\mathrm{d}n} \\ &= 1. \end{aligned} \tag{4}$$

This result, however, only holds when $n$ is large enough. What "large enough" means, however, is determined by the variance of the random variables $X_i$. The formal distinction is whether the distribution is "light-tailed" or "heavy-tailed", where the difference depends on whether the tail (the survival function) of the common distribution of the variables $X_i$ decreases faster (light-tail) or slower (heavy-tail) than an exponential tail. When $X_i$ are lognormally distributed they are heavy-tailed. Thus, $d_n = \mu n$ may not hold due to the large variance, except in the limit of sizes so large that are never attained in practice.

Can we find a sequence $d_n$ for lognormal distributions in a regime of "small sizes" (as opposed to "in the limit of extremely large sizes")?

Since lognormals with large variance tend to generate extremely large values, our approach will be to approximate $Y(n)$ by $\max\{X_1, \dots, X_n\}$. Of course, such approximation will be wrong for (extremely) large sizes $n$, where the scaling of $Y(n)$ is defined by the CLT. However, we will offer analytic justification for our heuristic approach, and we will find a sequence $d_n$ that we can use to characterize the sum $Y(n)$ for "small" $n$.

## *The maximum of lognormal random variables*

Let us write the lognormal random variable $X_i$ in the form $X_i = e^{\sigma Z_i + \ln x_0}$, where $Z_1, Z_2, \dots$ are i.i.d. random variables sampled from the standard normal distribution $\mathcal{N}(0,1)$. The important parameter in our analysis will be $\sigma$, the standard deviation of $\ln(X_i)$. We will adjust the other parameter, $x_0$, such that $\ln(x_0) = -\sigma^2/2$, in order to guarantee that the mean of the distribution is a constant $E[X_1] = e^{\ln(x_0)+\sigma^2/2} = 1$, chosen to be 1 for the purpose of convenience.

Our idea is to approximate $Y(n)$ by the maximal productivity $M(n) := \max\{X_1, \dots, X_n\}$ in the city. This quantity can be written as $M(n) = e^{\sigma L(n) - \sigma^2/2}$, where $L(n) := \max\{Z_1, \dots, Z_n\}$ denotes the maximum of i.i.d. standard normal random variables. The sum can be factorized such that $Y(n) = M(n)\big(1 + \epsilon_n(\sigma)\big)$, where the term $\epsilon_n(\sigma)$ is a series in which the dominant term (corresponding to the second largest value among $X_i$ divided by $M(n)$) is of order $e^{-\sigma/\sqrt{\ln(n)}}$. For $\sigma \gg \sqrt{ln(n)}$, the value of $\epsilon_n(\sigma)$ becomes negligible. In other words, because the parameter that determines the variance of lognormal random variables is very large compared to the size of the sample, the maximum dominates the sum. In Supplementary Information A we detail such analytical validation of the assumption that $Y(n)$ can be approximated by $M(n)$, and how much the other terms contribute to the sum.

The behavior of $L(n)$ for large $n$ is well-known (Leadbetter et al. 1983; Embrechts et al. 2013):

$$\lim_{n\to\infty} Pr\left(\sqrt{2\ln(n)}\left(L(n) - \sqrt{2\ln(n)} + \frac{\ln(\ln(n)) + \ln(4\pi)}{\sqrt{8\ln(n)}}\right) \leq x\right) = e^{-e^{-x}}, \quad x \in \mathbb{R}, \tag{5}$$

where the limit is the standard Gumbel distribution function. Thus, $L(n)$ grows as $\sqrt{2\ln(n)}$ with random fluctuations $c_n$ of the main order $\sqrt{2\ln(n)}$. Putting everything together, the sequence that tells us how the maximum $M(n)$ scales with size is approximately

$$d_n \approx \exp\left\{\sigma\sqrt{2\ln(n)} - \frac{\sigma^2}{2}\right\}. \tag{6}$$

We can thus state that for any fixed $\sigma$ large enough and $n$ sufficiently large (so that $L(n) \approx \sqrt{2\ \ln\ (n)}$) but still of order at most $e^{\sigma^2}$, we have

$$\begin{aligned} \beta &= \frac{\mathrm{d}\ln(d_n)\ /\mathrm{d}n}{\mathrm{d}\ln(n)\ /\mathrm{d}n}, \\ &\approx \frac{\mathrm{d}\left(-\frac{\sigma^2}{2} + \sigma\sqrt{2\ln(n)}\right)/\mathrm{d}n}{\mathrm{d}\ln(n)\ /\mathrm{d}n}, \end{aligned} \tag{7}$$

which yields

$$\beta(n,\sigma) \approx \frac{\sigma}{\sqrt{2\ln(n)}}. \tag{8}$$

Given our null model, equation (8) provides us with the expectation of the local scaling exponent from a purely statistical effect in the neighborhood of a specific sample of size $n$. The values $\beta(n,\sigma) \leq 1$ represent the combinations of $\sigma$ and $n$ for which one would expect constant returns to scale. Specifically, it is the region where $n$ is large enough relative to σ that the law of large numbers (LLN) applies, and no IRS should be observed.

## *Scaling exponent from a cross-section of many cities*

Scaling exponents are essentially local, since they quantify a relative rate of change of total output with size. Therefore, this rate may be different for different sizes, e.g. as shown by equation (8). In urban scaling analysis, however, we often estimate empirically a global scaling exponent that represents an average elasticity across many sizes using ordinary least squares (OLS) estimates (see, however, Gomez-Lievano et al. 2012; Leitão et al. 2016). Since the scaling coefficient may have a dependence on size $n$, the linearity assumptions underlying OLS regressions used in urban scaling analysis may not hold, which in turn may generate a bias in the estimate. Assuming the artificial IRS described in the previous section is present in data, what would be the scaling exponent if we were to estimate it from a regression line of the logarithm of total output against population size across many cities?

In a simple linear regression model $E[Y|X] = f(X)$ where $f(x) = a + bx$, the coefficient $b$ of the relation can be expressed as the ratio $\frac{Cov[X,Y]}{Var[X]} = \frac{E[XY]-E[X]E[Y]}{Var[X]}$. Thus, to compute the traditional global scaling exponent, we represent population size as a random variable $N$. In our case, the scaling exponent would be

$$\beta_{ave}(n_{min},\sigma,\alpha) = \frac{E[\ln(N)\ln(Y(N))] - E[\ln(N)]E[\ln(Y(N))]}{Var[\ln(N)]}.$$

So far, we have analyzed the behavior of $Y(N)$ for two regimes. One when $N$ is very large relative to $\sigma$ where we get equation (4) (i.e., the constant returns to scale guaranteed by the LLN for large sizes and small variances), and the other when in $N$ is small relative to $\sigma$ where we get equation (8) (the increasing returns to scale generated by the maximum for small sizes and large variances). Thus, if we generate a collection of cities in our null model in which $N$ takes several values from a distribution, the output $Y$ in the largest cities will scale linearly with size, while in small cities it will scale superlinearly. In order to account for all possible elasticities generated by both equations (4) and (8), we simplify our analysis by restricting the behavior of $Y(N)$ to these two scales only. Thus, in order to compute $\beta_{ave}(n_{min}, \sigma, \alpha)$, we define the following piecewise function

$$E[\ln(Y(N)) | \ln(N)] = \begin{cases} \ln(N), & \text{if } \ln(N) \geq \dfrac{\sigma^2}{2} \\ -\dfrac{\sigma^2}{2} + \sigma\sqrt{2\ln(N)}, & \text{if } \ln(N) < \dfrac{\sigma^2}{2}, \end{cases} \tag{9}$$

which combines both regimes.

As observed for firm and city sizes (see Saichev et al. 2009), we assume that $N$ is Pareto distributed, with probability density function

$$p_{N(n; n_{min}, \alpha)} = \frac{\alpha}{n_{min}} \left(\frac{n}{n_{min}}\right)^{-\alpha-1}, \quad \text{for } n \geq n_{min}. \tag{10}$$

When $\alpha = 1$, the distribution is often referred to as “Zipf’s Law”. The parameter $n_{min}$ determines the minimum value above which sizes follow a Pareto distribution.

Using equation (9) and computing expectation values with equation (10), we can solve for the value of the average scaling exponent $\beta_{ave}(n_{min}, \sigma, \alpha)$. In the Supplementary Information C we detail the important steps of this derivation, which yields

$$\beta_{ave}(n_{min}, \sigma, \alpha) = 1, \quad \text{for } n_{min} \geq e^{\frac{\sigma^2}{2}}, \tag{11}$$

and

$$\beta_{ave}(n_{min}, \sigma, \alpha) = \frac{e^q\sqrt{\pi w}(1-2q)}{2}\left(\operatorname{erf}(\sqrt{w}) - \operatorname{erf}(\sqrt{q})\right) + \sqrt{wq} + e^{q-w}(1-q), \quad \text{for } n_{min} < e^{\sigma^2/2}, \tag{12}$$

where $q \equiv \alpha \ln(n_{min})$ and $w \equiv \frac{\alpha\sigma^2}{2}$. Equation (12) represents our main analytic contribution.

Equation (12) provides the null expectation for the scaling exponent of urban productivity with size, under the assumption of i.i.d. lognormal productivities and Pareto sizes. Notice that $\beta_{ave}(n_{min}, \sigma, \alpha)$ is a function of the parameters of the distributions only.

Research is often done on random subsamples of workers across cities. We model this situation introducing a new parameter $f$, between 0 and 1, pre-multiplying all populations. Conveniently, the change of variable $n' = fn$ does not change probabilities of events $p_N(n)\mathrm{d}n = p_{N'}(n')\mathrm{d}n'$ (with $n'_{min} = f\, n'_{min}$) based on the Pareto density in equation (10). This change only affects the parameter $n_{min}$, so we have $\beta_{ave}(f\, n_{min}, \sigma, \alpha)$. For example, if one is working with a random 1% census sample, it suffices to multiply the parameter $n_{min}$ in equation (12) by $f = 0.01$.

Figure 1 shows three graphs, plotting $\beta_{ave}(f\, n_{min}, \sigma, \alpha)$ as a function of one of the parameters, keeping the other parameters fixed. Panel A shows constant returns to scale (i.e., $\beta = 1$) for small $\sigma$, but increasing returns ($\beta > 1$) for large $\sigma$. Panel B confirms the effect of the LLN: larger percentages of the city populations reduce the artificial IRS. Finally, Panel C shows that $\beta$ increases with $\alpha$, which suggests that the scaling exponent will be larger the more uniform are city sizes.

Of all three parameters, $\alpha$ has the weakest effect on $\beta$. Its effect in practice is probably negligible given the fact the estimated values from real data barely deviate from $\hat{\alpha} \approx 1$. In contrast, parameters $\sigma$ and $f$ strongly affect the values of $\beta$. In the following sections we will analyze the effect of $\sigma$ through simulations, and the effect of $f$ with real-world data.

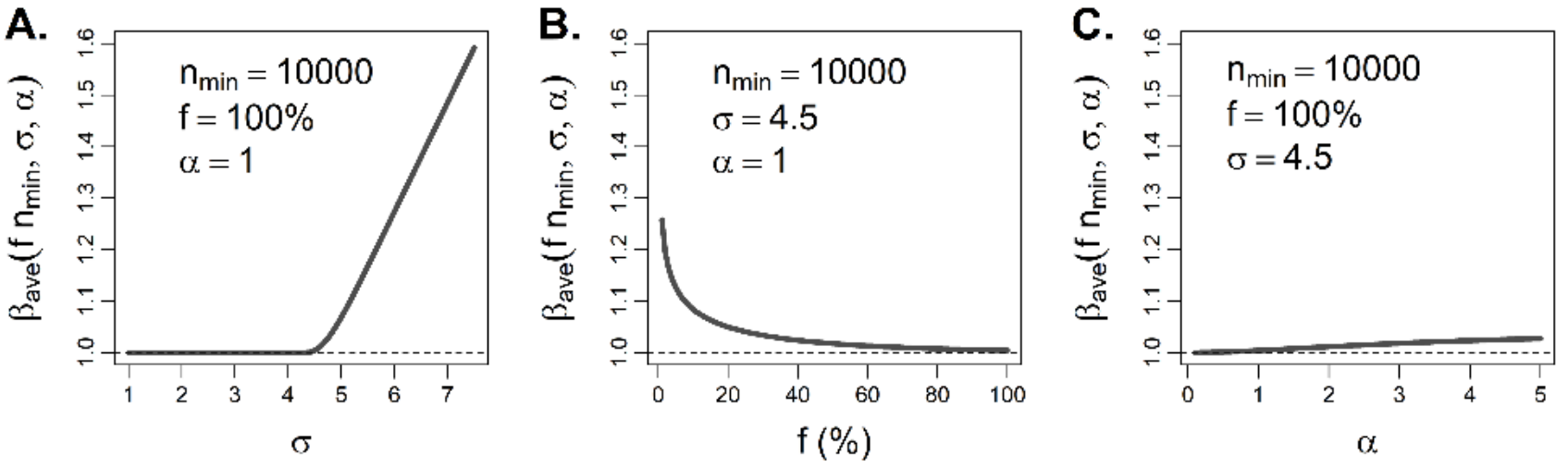

**Figure 1.** Artificial scaling exponent of total urban output with respect to city size, as a function of three distributional parameters. Panel **A**: Dependence on $\boldsymbol{\sigma}$, the standard deviation of log-productivities. Panel **B**: Dependence on ***f***, the fraction sampled from city population sizes. Panel **C**: Dependence on $\boldsymbol{\alpha}$, the Pareto coefficient for the distribution of city sizes.

## Simulations

We simulate $m = 900$ synthetic cities using our model. Each city has a population $n_k$ taken from a Pareto distribution with parameters $n_{min} = 10{,}000$ and $\alpha = 1$. For each city $k$ we generate $n_k$ productivities sampled i.i.d. from equation (1). We will make use of all individuals, so $f = 1$. We will generate simulations for different values of the model parameter $\sigma$, and we will compare $y^{(s)}(n_k)/n_k$ against $n_k$, for all cities $k = 1, \dots, 900$, where we will use the superscript (s) to make explicit the fact that the output of cities is simulated.

Figure 2 shows the results of such simulations, plotting per capita productivity with respect to population size, using logarithmic axes. The black dashed line is the theoretical expected value of average productivity, which we set to $\mu = 1$. The simulated data in figure 2 are well described by the relation $\frac{y(n)}{n} = Y_0 n^{\delta}$. The purple solid line is the OLS fit of the typical urban scaling relationship

$$\ln\left(\frac{y^{(s)}(n_k)}{n_k}\right) = \ln(Y_0) + \delta \ln(n_k) + \varepsilon_k$$

where the estimated values of $\delta$, as we vary the value of σ in our simulations,

should be compared to predicted values of $\delta_{ave}(f\, n_{min}, \sigma, \alpha) = \beta_{ave}(f\, n_{min}, \sigma, \alpha) - 1$. Therefore, the total output of the city under our model is well approximated by the power-law function

$$F_{model}(n) = Y_0 n^{\beta}, \tag{13}$$

where $\beta = 1 + \delta$.

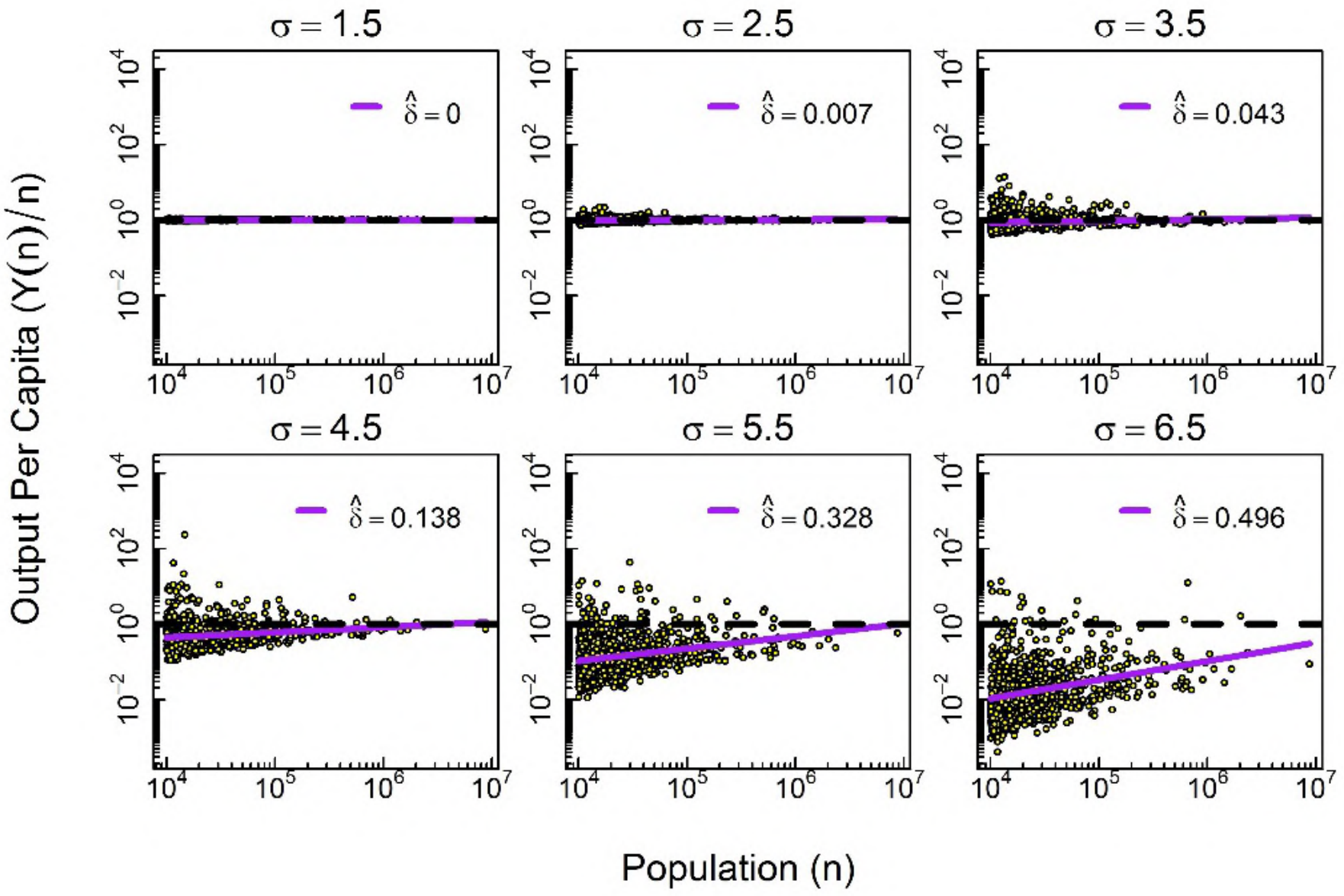

**Figure 2.** Effect of $\boldsymbol{\sigma}$ on output per capita in null model. Dots represent $\boldsymbol{m}$ = 900 cities with populations generated from a Pareto distribution of parameters $\boldsymbol{n}_{\min}$ = 10 000 and $\boldsymbol{\alpha}$ = 1. Individual lognormal productivities are generated such that $E[X] = 1$ is fixed (black dashed line). Ordinary Least Square (OLS) regression line of $\ln(\boldsymbol{Y}(\boldsymbol{n})/\boldsymbol{n})$ against $\ln(\boldsymbol{n})$ is shown as the purple solid line.

As can be observed, the parameter $\sigma$ controls the artificial scaling exponent $\beta$ exactly as we anticipated from our analytic predictions. We also note that the estimated $Y_0$ decreases as $\sigma$ increases. That all these effects emerge when the variance of the log-productivity is very large is also reflected on the fact that the goodness-of-fit of equation (13) decreases, as evidenced by the low $R^2$ values in Figure 3C below.

Figure 3 plots more systematically the departure of $\hat{\beta}$ and $\widehat{\ln(Y_0)}$ from their theoretical values, $\beta = 1$ and $\ln(Y_0) = \ln(\mu) = 0$, and their dependence on $\sigma$. Clearly, an urban scaling law emerges as σ increases. The plot was constructed by simulating 100 different runs of the model (i.e., 100 different cross-sections of $m$ cities defined by the ordered pairs $(n_k, y^{(s)}(n_k))$) per each value of $\sigma$ between 1.5 and 6.5. We observe that the value of $\beta$ starts to depart from 1.0 when $\sigma \approx 3.0$, qualitatively following the predictions from equation (12) and figure 1. In panel A of figure 3, the dotted line represents the analytical curve predicted by $\beta_{ave}(n_{min}, \sigma, \alpha)$ for $n_{min} = 10{,}000$ and $\alpha = 1.0$. It is important to note that, for each value $\sigma$, the gray area representing the region where 95% of estimated values of $\beta$ across simulations is relatively narrow, which means that the average scaling exponent departs from 1, *significantly and systematically*. These departures from the theoretical values are associated with a larger unexplained variance of the OLS regression, which we observe as a monotonically decreasing $R^2$.

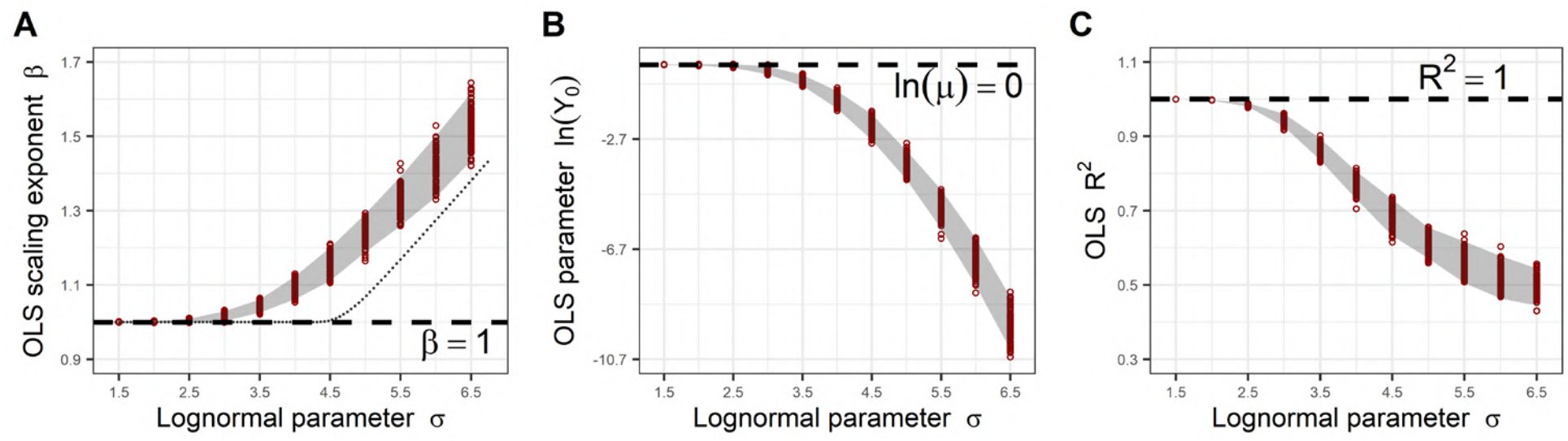

**Figure 3.** Artificial scaling exponents driven by a departure from the Law of Large Numbers (LLN) as lognormal productivities become more heavy-tailed. Each point represents an estimate of a linear regression for a single cross-section of cities, like one of the panels in figure 2, for which we show the OLS estimations of: the scaling exponent $\boldsymbol{\beta}$ (panel **A**), the intercept $\ln(\boldsymbol{Y_0})$ (panel **B**), and the $\boldsymbol{R^2}$ of the regression (panel **C**). For each value of $\boldsymbol{\sigma}$ we generated 100 simulations. The gray areas show the regions where 95% of the point estimates fell. The dotted line in panel **A** is the scaling exponent predicted by Equation (12).

In the next section we will analyze the effects of decreasing $f$, while keeping $\sigma$ fixed. The prediction is that the city size premium will artificially become larger with increasingly smaller sample sizes (see panel B of figure 1). For this, we will analyze real data on Colombian wages.

## An Application

Equation (8) highlights two important effects. On the one hand, that the artificial scaling exponent $\beta$ of total output in a city with respect to its population size will increase if the standard deviation of log-productivity, $\sigma$, increases. On the other, it tells us that $\beta$ will also increase when sample sizes $n$ (or, rather, $fn$) are small. The former prediction was analyzed in the last section through simulations. The latter prediction is studied in this section.

### *Data, Descriptives and Distributions*

The data used here is the 2014 administrative records of the social security system in Colombia (the Spanish acronym is PILA, for Integrated Report of Social Security Contributions) to analyze the average monthly wages across formal workers in all Colombian municipalities. We refer the reader to the Supplementary Information B for the source, and details about the cleaning and preparation of the dataset. After the preprocessing of the data, the final sample consists of a total of 6,713,975 workers employed in the formal sector, geographically distributed in 1,117 municipalities that cover almost the entire Colombian territory. We quantify the "population size" of a municipality using the count of formal employment, defined as the number of workers in our data that reported the municipality as the last place of work in 2014.

We first want to assess whether a lognormal and a Pareto characterize the distribution of wages and sizes, respectively. In Supplementary Information D we present a detailed analysis of the goodness-of-fit statistics for several probability distribution functions to model wages (Table 1) and municipality sizes (Table 2). Both quantities are left-truncated, so we fitted some truncated probability distribution functions through Maximum Likelihood Methods. Wages were best fit by a truncated-lognormal, and municipality sizes were best fit by a Pareto distribution, according to three criteria: the largest likelihood, the minimum Akaike Information Criterion (AIC), and the minimum Bayesian Information Criterion (BIC).

The fitted distributions yielded the following estimated parameters: $\hat{\sigma} \approx 2.00$, $\hat{\alpha} \approx 0.67$ and $\widehat{n_{min}} \approx 287$ (see Supplementary Information D for graphical comparisons and estimated confidence intervals of these parameters, figures 7 and 9), which we can use to anticipate whether to expect an artificial IRS.

Our results state that as we take smaller random fractions $f$ of the total population, we should observe the artificial appearance of a superlinear scaling exponent if the condition $f n_{min} < \mathrm{e}^{\sigma^2/2}$ approximately holds. For Colombian estimates, this means we will observe an artificial scaling exponent for $f < 0.026$. This implies taking seven workers at random from the smallest municipality, and up to $58{,}495$ workers from the largest municipality. We will use equation (12) to anticipate how $\beta_{ave}(f\ 287, 2, 0.67) > 1$ as we reduce sizes using fractions of the data for $f$ less than 0.026. We will study the effect of taking small samples of workers in more detail below.

## *Telling Apart Real Versus Artificial Scaling Exponents*

Figure 4 plots the cross-section of the average monthly wage per municipality with respect to municipality size. There is clearly a positive and significant scaling exponent $\hat{\delta} \approx 0.06$.

The strategy to identify whether a scaling exponent of wages with respect to size is due to an artificial sampling effect is to randomize workers geographically. The reasoning behind this is fairly clear: while randomizing individuals should eliminate the empirical evidence for urban productivity premiums given by the built-in dependencies of individuals caused by sorting or agglomeration effects, the artificial IRS effect should be statistically invariant to the removal of the causal effects present in the data. Randomizing will destroy the information of the way workers have sorted themselves across cities, and of who the workers have interacted, or are interacting, with. In other words, the causal effects are removed by randomizing the spatial location of workers, but the distributional effects are not. After we randomize the municipalities where workers work, any scaling exponent remaining from a regression must come from the statistical sampling effect of the distribution.

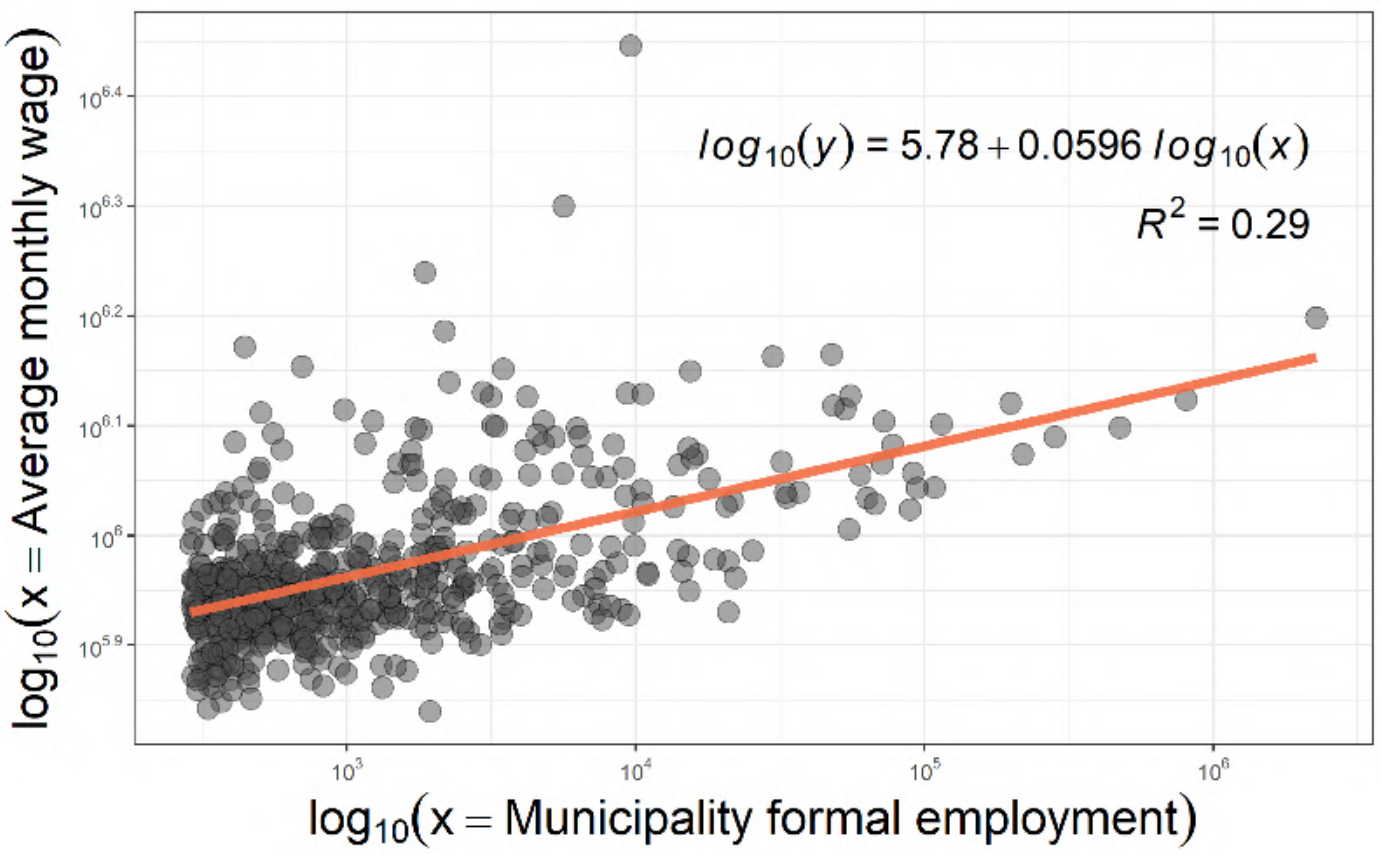

**Figure 4.** Data from Colombian formal workers in 2014 show that larger municipalities have, on average, workers with higher monthly wages.

Notice that randomization does not change workers' wages. In this sense, we have not destroyed *all* of the information, since the distribution of wages is itself a consequence of the socioeconomic causes related to people moving, agglomerating, and learning from each other. Hence, we are not claiming that the geographical randomization of people assumes workers would have earned that same wage had they worked in that new location. We are also not claiming that the distribution of wages should be invariant to the presence or absence of sorting or agglomeration effects. We are just saying that *conditional on that distribution*, scaling exponents larger than one can arise naturally and systematically in a regression, even after destroying the local information attached to where people are located.

For the real and the randomized versions of the data, we will estimate the following basic regression:

$$\ln\left(\overline{w_k}^{(f,j)}\right) = \alpha + \delta \ln(n_k) + \varepsilon_k, \tag{14}$$

where $k$ is the index for municipality $k$, and $j \in \{real, randomized\}$, where "*real*" indicates that we compute the average wage from the actual individuals that work in municipality $k$, whereas "*randomized*" indicates that we are taking the average after randomly permuting the location of individuals across municipalities. The superscript $f$ is to indicate that the average wage (real or randomized) was taken over a fraction $f$ of all workers. We will explore the effect on scaling exponents from taking different values of $f$. In the regression given by equation (14), however, the size of formal employment $n_k$ for each municipality is kept as the total count of formal employment, and does not vary with $f$ or $j$.

The method is to (i) take a subsample $f$ from the full population of workers, (ii) compute their average wage to estimate the ("real") scaling exponent of wages with respect to employment size, (iii) randomize the location of individuals, (iv) compute the new average wage per municipality and estimate the resulting ("randomized") scaling exponent. For a given subsample $f$, we do (iii) and (iv) 1,000 times. Furthermore, we repeat this whole process, (i)-(iv), 10 times so that we can compare different subsamples determined by the same $f$. We obtain form this process a distribution of possible scaling exponents from sampling effects.

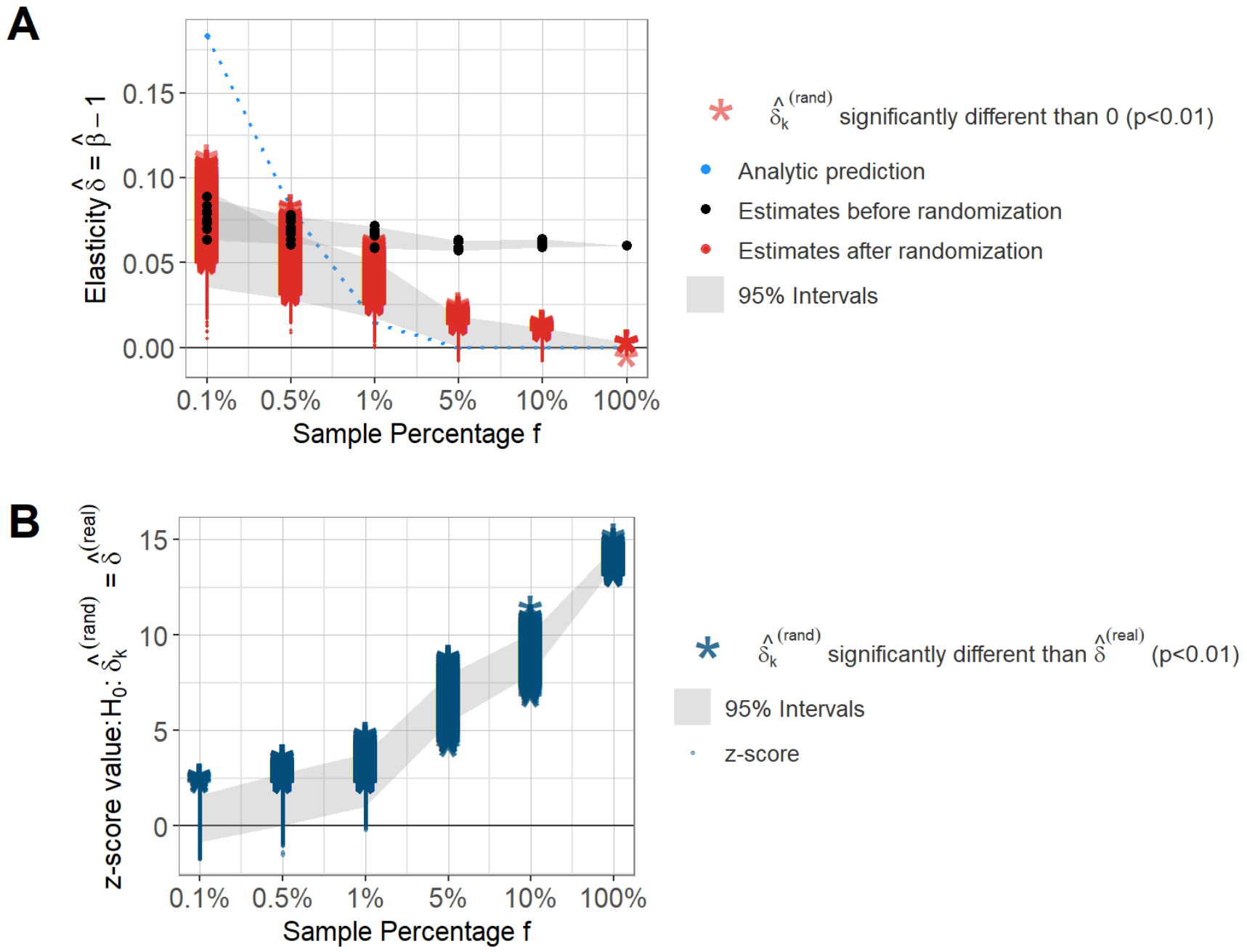

**Figure 5.** Effects on scaling exponents from decreasing sample sizes by reducing the number of sampled individuals per city. Panel **A** plots the elasticities (y-axis) calculated before randomizations (black dots) and after randomizations (red dots), for a given subsample of workers determined by each of the values of parameter ***f*** (x-axis). It is observed that as ***f*** decreases, elasticities increase (see main text for details about the procedure). Those red dots that are statistically different from zero have been highlighted by a red star. The dotted blue line is the scaling exponent $\boldsymbol{\delta} = \boldsymbol{\beta_{ave}(f\, n_{min}, \sigma, \alpha) - 1}$ predicted by Equation (12), with values $\boldsymbol{n_{min} = 287}$, $\boldsymbol{\sigma = 2.0}$, $\boldsymbol{\alpha = 0.67}$, as a function of $\boldsymbol{f}$. Panel **B** plots the values of the ***z***-score statistic for each scaling exponent from the OLS regression after individuals have been randomized, constructed in order to test the null hypothesis that, given a subsample of the workers, the scaling exponent after individuals have been randomized is equal to scaling exponent before randomization. Scaling exponents of the randomized samples that are statistically different from the corresponding scaling exponent without randomization have been highlighted by a blue star.

The null hypothesis to be tested in this procedure is that elasticities without randomization are equal to the elasticities after randomization. Notice that if the LLN holds for our sample, elasticities after randomization should be zero, which is the specific case in which most urban scaling analysis has been carried out.

We assume a specific subsample of workers, and let $\hat{\delta}^{(real)}$ be the estimated elasticity without the randomization, and $\hat{\delta}_l^{(rand)}$ the elasticity after one specific $l$-th randomization. Since these regression coefficients are OLS estimates, a path to move forward into creating a test statistic is to assume these estimates follow a normal distribution, and have a standard error associated with them, $se^{(real)}$ and $se_l^{(rand)}$, respectively. Under the null hypothesis that these two estimated elasticities are equal, and assuming the number of municipalities large, we can construct the following $z$-score (see Clogg et al. 1995; Paternoster et al. 1998):

$$z_{\text{stat}} = \frac{\hat{\beta}^{(real)} - \hat{\beta}_l^{(rand)}}{\sqrt{(se^{(real)})^2 + \left(se_l^{(rand)}\right)^2}}, \tag{15}$$

which will follow approximately a standard normal distribution.

Figure 5, panel A, plots the elasticities before and after randomizations (black and red dots, respectively). Those elasticities after randomization that are statistically significant (at a level $p < 0.01$) have been highlighted with a red large star marker "*". Since for each subsample we generate $1{,}000$ randomizations, we also show the bands between which $95\%$ of the elasticities fall. The blue dotted line in panel A shows our analytic prediction. As can be observed, we confirm that the elasticities after randomizing individuals decrease steadily as larger samples are taken. For $f < 0.01$, however, many elasticities are not significantly different anymore from the real ones. This is shown in panel B, which plots the $z$-score given by equation (15).

## Discussion and conclusions

We have presented extensive evidence that increasing returns to scale (IRS) implied by superlinear urban scaling can emerge in the total absence of self-sorting, externalities, or interactions. The main methodological implication from our work is that the null hypothesis in urban scaling analysis of wages should not necessarily be the absence of IRS (i.e., $\beta = 1$), since IRS ($\beta > 1$) can be observed under certain conditions even when the data generating process does not have the putative underlying mechanisms.

In our analytical results, we showed that the elasticity emerging from the effect we presented here depends positively on the standard deviation of log-productivities, and negatively on the sample sizes. We derived a precise formula to compute the null expected elasticity for a single city and for a cross-section with different sizes. The scaling exponent on the cross-section becomes solely a function of the distributional parameters of productivities and sample sizes. Our approach shifts attention away from the study of averages to the analysis of probability distributions (see, e.g., Gould 1996; Gabaix 2009; Mantovani et al. 2011; Gomez-Lievano et al. 2012; Behrens and Robert-Nicoud 2015; Leitão et al. 2016).

Given access to the full population of Colombian formal workers in 2014, we illustrated the statistical emergence of an artificial IRS in real data for random subsamples smaller than 1% of the total population of workers in our data. As predicted, we confirmed that the artificial urban scaling effectively disappears for samples larger than that.

In general, our present study highlights the importance of analyzing with care data from small samples, or surveys. One must understand the distributional properties that describe individuals, like how the variance relates to the possible sample sizes, before carrying out aggregations. This line of research warns about what increasing returns to scale might imply from a statistical point of view when measures of individual output are unevenly distributed. For example, the equivalence between "total output increases more than proportionately with size" and "individual productivity increases with larger sizes" is only applicable when the law of large numbers is valid, meaning that per capita transformations may give misleading information about actual average individual productivities.

We need to be alert for violations of the law of large numbers. When productivities or wages are fat-tailed, our null expectations should not be the statistical absence of a size effect, but rather the presence of it. We hope that further analysis of the effect of size on productivity and wages will account for these distributional effects.

## Acknowledgements

We gratefully acknowledge useful comments from Ricardo Hausmann, Luis M.A. Bettencourt, Rachata Muneepeerakul, Dario Diodato, and Michele Coscia. Special thanks go to Frank Neffke, James McNerney, and Ulrich Schetter, for many helpful comments and suggestions, which improved the clarity in the presentation of some of the central results in this project. This research did not receive any specific grant from funding agencies in the public, commercial, or not-for-profit sectors. Declarations of interest: none.

# Supplementary Information
# Artificial Increasing Returns to Scale and the Problem of Sampling from Lognormals

## *Supplementary Information A. Validating the assumption that the sum $Y(n)$ can be approximated by the maximum $M(n)$*

We approximate $Y(n)$ by the maximal productivity $M(n) := \max\{X_1, \ldots, X_n\}$ in the city. This quantity can be written as $M(n) = e^{\sigma L(n) - \sigma^2/2}$, where $L(n) := \max\{Z_1, \ldots, Z_n\}$ denotes the maximum of i.i.d. standard normal random variables. Then

$$\begin{aligned} Y(n) &= \sum_{i=1}^{n} X_i \\ &= \sum_{i=1}^{n} \mathrm{e}^{\sigma Z_i - \sigma^2/2} \\ &= \mathrm{e}^{\sigma L(n) - \sigma^2/2} \sum_{i=1}^{n} \mathrm{e}^{\sigma (Z_i - L(n))} \\ &= M(n) \sum_{i=1}^{n} \mathrm{e}^{\sigma (Z_i - L(n))}. \end{aligned} \tag{16}$$

The main difficulty for validating the assumption that $Y(n)$ can be approximated by $M(n)$ is in analyzing the last sum in Equation (16). Since it is doubtful that this quantity can be tackled analytically, we suggest the following argument. First write

$$\Delta_n := \sum_{i=1}^{n} e^{\sigma (Z_i - L(n))} = \sum_{i=1}^{n} \mathrm{e}^{\sigma (L_i(n) - L(n))},$$

where we have re-ordered the terms in the summation such that $L_i(n)$ denotes the $i$th largest value among $Z_1, \ldots, Z_n$. For the first term, we have $L_1(n) = L(n)$, so $\mathrm{e}^{\sigma (L_1(n) - L(n))} = 1$. For the second term, we can use that $L(n) - L_2(n)$ is of order $(\ln(n))^{-1/2}$ (see Leadbetter et al. 1983, Section 2.3). By our assumption that $\sigma \gg \sqrt{2\ln(n)}$, the quantity $\sigma(L_2(n) - L(n))$ is negatively large and so $\mathrm{e}^{\sigma(L_2(n) - L(n))}$ is close to 0. The remaining terms $\mathrm{e}^{\sigma(L_i(n) - L(n))}$ for $i \geq 3$ decay to 0 much faster since so do exponents with larger negative powers.

Thus, we have $\Delta_n \approx 1$ when $\sigma \gg \sqrt{\ln(n)}$. Moreover, our simulations (not shown) reveal that a similar conclusion applies even when $\sigma$ is larger than, but comparable to, $\sqrt{2\ln(n)}$, in which case $\Delta_n$ is rather close to 1, being of constant order.

Figure 6 illustrates the fact that the maximum can indeed become comparable to the sum. We use a proxy of the share $M(n)/Y(n)$ as the ratio of the quantile $Q(1 - 1/n)$ over $\sum_i Q(i/n - 1/n)$, where $Q(\Pr(X \leq x_p)) = x_p$. The figure shows the curves for this proxy of the share $M(n)/Y(n)$ as a function of $\sigma$, for four distinct values of $n$.

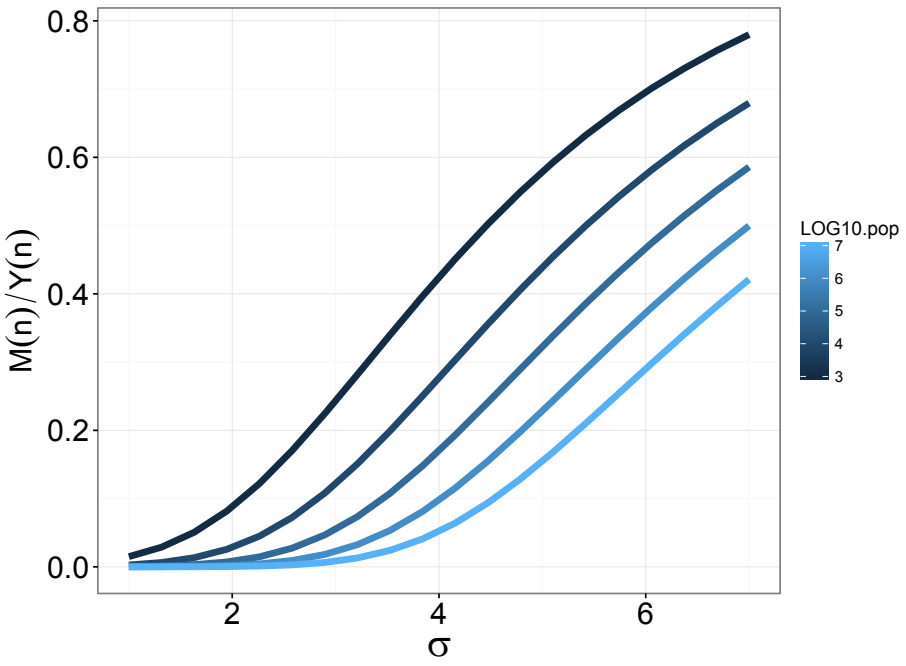

**Figure 6.** Proxy for the share of the maximum over the total sum, $M(n)/Y(n)$, constructed as the ratio of the quantile of the lognormal associated with the percentile $(n-1)/n$ over the sum of all the $n$ quantiles, as a function of parameter $\sigma$. The color of the line represents a fixed population size. We show the curves for $n = 10^3$, $10^4$, $10^5$, $10^6$, and $10^7$. The lighter the blue is, the larger the population is.

According to our derivations, for $\sigma = 4$, the maximum is comparable to the sum when $n < \mathrm{e}^{\sigma^2/2} \approx 3,000$. Indeed, the figure shows that for $\sigma = 4$ and for $n = 10^3$ (darkest blue line), the maximum can account for $50\%$ of the sum. For $\sigma = 4$ one needs to increase size to $n = 10^7$ (a ten-thousand-fold increase) in order to decrease the dominance of the maximum to about $10\%$ (see lightest blue line). Figure 6 provides evidence in support of replacing the sum $Y(n)$ with the maximum $M(n)$, and using this to find an approximate result for how the sum scales with size.

## *Supplementary Information B. Data*

In the main text we use data of the formal workforce in Colombia to analyze the unconditional elasticity of nominal wages on municipality population size. These come from the administrative records of the social security system in Colombia (abbreviated as PILA in Spanish, meaning the *Integrated Report of Social Security Contributions*). The PILA is maintained by the Colombia Ministry of Finance and Public Credit ("Ministerio de Hacienda y Crédito Público"). PILA consists of individual contributions to health and pensions reported by workers, firms, public institutions, and other formal entities like associations, universities, cooperatives and multilateral organizations.

The dataset was obtained from the Ministry of Finance and Public Credit, under a data use agreement that is part of the development of `www.datlascolombia.com`, a joint project between the Center for International Development and the Colombian Foreign Trade Bank (Bancoldex) to map the industrial economic activity in Colombia. The data are stored on secure computers at the Harvard-MIT Data Center. Access is restricted to identified and authorized researchers by means of a confidential account. The use of the PILA for research purposes has been reviewed by the Harvard's Institutional Review Board (IRB). In the database individuals and firms have been previously anonymized in order to protect their habeas data. Harvard IRB determined that this dataset is not human subjects as defined by the Department of Health and Human Services (DHHS) regulations.

Each row of the dataset consists of a monthly contribution to the social security system, with more than seventy different fields with information about the worker and the firm, and with the values of the contribution to health and pension, according to the days the worker worked at the firm in that month. The raw microdata consists of 122,287,562 rows (i.e., social security contributions), from 10,535,587 unique workers (i.e., each worker had an average of 11.6 contributions per year). As explained below, we aggregate and keep a subset of all these observations, and we only use two fields for this study: the list of nominal wages earned, and the municipalities of work to which the wage values where attached.

As a start, these data must be cleaned, as is often the case with datasets built from observations resulting from administrative transactions. Common problems include misreported or missing wages, no municipality of work reported, no age reported, duplicated observations, or missing contribution to pension or health. In addition to dropping these problematic observations, we keep only those workers that are categorized as "dependent" or "independent", which means they are either employed in a firm or are self-employed, respectively (by keeping these type of social security contributors we exclude those individuals that contribute to social security through means other than a formal job). Finally, we keep those individuals who worked for at least 30 days during the whole year, and had ages between 15 and 64.

We compute the monthly average wage of workers by first adding their net wage earned during the year, then dividing it by the total number of days worked, and finally multiplying by thirty. By law, firms are required to pay a minimum wage to workers, or more. However, there exist special cases in the dataset in which this does not hold. Hence, we make sure this is the case by dropping observations which report average monthly wages below the minimum wage ($616,000 Colombian Pesos, or COP, in 2014). At the end, our population of analysis consists of 6,713,975 formal Colombian workers (approximately 64% of the unique individuals that appear originally in the dataset).

## *Supplementary Information C. Derivation of* $\beta_{ave}(n_{\min}, \sigma, \alpha)$

Here we shall indicate the relevant steps for the derivation of Equation (12).

The derivation becomes relatively easy once some changes of variables are carried out first. We start with the change of variable $U = \alpha \ln(N)$. Together with equation (10) where $N \sim Pareto(n_{\min}, \alpha)$, and the conservation of probability, we get that $p(u) = p(n) \left|\frac{\mathrm{d}u}{\mathrm{d}n}\right|^{-1} = \mathrm{e}^{-(u-q)}$, for $u \geq q$, where $q \equiv \alpha \ln(n_{\min})$. That is, a shifted standard exponential,

$$U \sim q + Exp(1).$$

With the additional change of variable $V = \ln(Y)$, the piecewise function in equation (9) can be re-written as

$$\mathrm{E}\left[V|U\right] = \begin{cases} \frac{\sigma^2 U}{2w} & , \text{ for } U \geq w \\ -\frac{\sigma^2}{2} + \frac{\sigma^2}{\sqrt{w}}\sqrt{U} & , \text{ for } U < w, \end{cases}$$

where $w \equiv \alpha\sigma^2/2$. Using the indicator function, we express the equation above more concisely as

$$\mathrm{E}\left[V|U\right] = \frac{\sigma^2}{2w}\left[U + \left(2w^{1/2}U^{1/2} - w - U\right)\mathbb{1}_{\{U<w\}}\right].$$

For the computation of $\beta_{ave}(n_{\min}, \sigma, \alpha)$ we have to get analytic expressions of the different terms in the following relation (see main text):

$$\begin{aligned}\beta_{ave}(n_{\min}, \sigma, \alpha) &= \frac{\mathrm{E}\left[(U/\alpha)V\right] - \mathrm{E}\left[(U/\alpha)\right]\mathrm{E}\left[V\right]}{\mathrm{Var}\left[(U/\alpha)\right]} \\ &= \alpha\left(\frac{\mathrm{E}\left[UV\right] - \mathrm{E}\left[U\right]\mathrm{E}\left[V\right]}{\mathrm{Var}\left[U\right]}\right).\end{aligned}$$

The easy terms are those in which $U$ is alone:

$$\begin{aligned}\mathrm{E}\left[U\right] &= 1 + q, \\ \mathrm{Var}\left[U\right] &= 1.\end{aligned}$$

The piecewise form of $\mathrm{E}\left[V|U\right]$, however, complicates the rest. Let us compute $\mathrm{E}\left[V\right]$ using the law of total expectations $\mathrm{E}\left[V\right] = \mathrm{E}\left[\mathrm{E}\left[V|U\right]\right]$:

$$\mathrm{E}\left[V\right] = \int_q^\infty \frac{\sigma^2}{2w}\left[u + \left(2w^{1/2}u^{1/2} - w - u\right)\mathbb{1}_{\{u<w\}}\right]p(u)\mathrm{d}u.$$

The first term is trivial, since it is simply the expectation of $U$. For the second term, since the integral starts at $q$, the indicator function $\mathbb{1}_{\{u<w\}}$ can only be true whenever $w$ is also larger than $q$. Hence, we can compute the second term in the integral by taking the indicator function out of the integral and replacing it with $\mathbb{1}_{\{q<w\}}$, and evaluating the integral only between $q$ and $w$:

$$\mathrm{E}\left[V\right] = \frac{\sigma^2}{2w}\left[(1+q) + \mathbb{1}_{\{q<w\}}\int_q^w \left(2w^{1/2}u^{1/2} - w - u\right)p(u)\mathrm{d}u\right].$$

An almost identical expression can be derived for $\mathrm{E}\left[UV\right]$.

Now, given that $p(u) = \mathrm{e}^q\mathrm{e}^{-u}$, one recognizes that most terms become integrals of the form

$$\int_q^w u^s\mathrm{e}^{-u}\mathrm{d}u = \gamma(s+1, w) - \gamma(s+1, q),$$

for some constant $s$, where we use "lower incomplete gamma functions", defined as $\gamma(s+1, x) = \int_0^x u^s\mathrm{e}^{-u}\mathrm{d}u$. We use the recurrence relationship

$$\gamma(s+1, x) = s\gamma(s, x) - x^s\mathrm{e}^{-x}$$

iteratively, as many times as necessary until we either get to terms like $\gamma(1/2, x)$, or terms like $\gamma(1, x)$. For the former, we use the fact that $\sqrt{\pi}\mathrm{erf}(\sqrt{x}) = \gamma(1/2, x)$, to express everything in terms of "error functions", defined as $\mathrm{erf}(x) = (2/\sqrt{\pi}) \int_0^x \mathrm{e}^{-t^2}\,\mathrm{d}t$. For the latter, we use the (trivial) fact that $\gamma(1,x) = 1 - \mathrm{e}^{-x}$.

After reducing the relevant integrals of $\mathrm{E}\left[V\right]$ and $\mathrm{E}\left[UV\right]$ to simple exponentials and error functions, it is just a matter of collecting terms. Finally, we arrive to the final expression:

$$\beta_{ave}(n_{\min},\sigma,\alpha) = \begin{cases} 1, \quad \text{for } q \geq w \\ \\ \mathrm{e}^{q-w}(1-q) + (wq)^{1/2} \\ \qquad +\frac{\mathrm{e}^{q}(\pi w)^{1/2}}{2}(1-2q)\left(\mathrm{erf}(w^{1/2}) - \mathrm{erf}(q^{1/2})\right), \quad \text{for } q < w. \end{cases}$$

With some minor replacements, the reader can check that this is the same as equation (12) in the main text.

## *Supplementary Information D. Tables for goodness-of-fit statistics for monthly wages and municipality sizes in Colombia*

Here, we show some tables comparing different alternative distributions and their goodness-of-fit for wages and sizes, and we show some comparative graphs.

### *Wages*

**Table 1.** Distributions fitted to individual wages. The number of total workers ($1,325,950$ observations) analyzed in this table differ from the number mentioned in the main text ($6,633,449$) because wages are clustered on the minimum wage. The fits of continuous distributions to data with repeated values, such as the minimum value which is repeated several times, was much improved when we removed repeated values. The list of the distributions are ordered from top to bottom by increasing AIC values.

| dist | numobs | loglik | AIC | BIC | Parameter 1 | C.I. | Parameter 2 | C.I. |
|---|---|---|---|---|---|---|---|---|
| trunclnorm | $1,325,950$ | -$19,940,740$ | $39,881,484$ | $39,881,508$ | $\widehat{\ln(x_0)} = 10.23$ | [2, 10.15] | $\widehat{\sigma} = 2.00$ | [1.99, 2.02] |
| powerlaw | $1,325,950$ | -$19,952,047$ | $39,904,095$ | $39,904,108$ | $\widehat{\alpha} = 1.16$ | [1.16, 1.16] | | |
| trunccauchy | $1,325,950$ | -$19,960,472$ | $39,920,948$ | $39,920,972$ | $\widehat{l} = 120,766.20$ | [190129.96, 114817.91] | $\widehat{s} = 190,130.00$ | [174740.73, 206786.87] |
| truncgamma | $1,325,950$ | -$20,018,427$ | $40,036,859$ | $40,036,883$ | $\widehat{a} = 0.0000$ | [0, 0] | $\widehat{\lambda} = 0.0000$ | [0, 0] |
| truncweibull | $1,325,950$ | -$20,077,580$ | $40,155,164$ | $40,155,188$ | $\widehat{a} = 0.72$ | [0.72, 1109695.37] | $\widehat{b} = 1.11 \times 10^6$ | [1109654.53, 1111878.12] |
| truncgumbel | $1,325,950$ | -$20,330,914$ | $40,661,832$ | $40,661,856$ | $\widehat{a} = 0.13$ | [0.13, 1332540.63] | $\widehat{b} = 1.33 \times 10^6$ | [1330795.1, 1332942.21] |
| lnorm | $1,325,950$ | -$20,344,669$ | $40,689,342$ | $40,689,366$ | $\widehat{\ln(x_0)} = 14.19$ | [0.76, 14.19] | $\widehat{\sigma} = 0.76$ | [0.76, 0.76] |
| gamma | $1,325,950$ | -$20,626,561$ | $41,253,126$ | $41,253,151$ | $\widehat{a} = 1.41$ | [0, 1.41] | $\widehat{\lambda} = 0.0000$ | [0, 0] |
| weibull | $1,325,950$ | -$20,667,064$ | $41,334,133$ | $41,334,157$ | $\widehat{a} = 1.05$ | [1.05, 2219332.54] | $\widehat{b} = 2.22 \times 10^6$ | [2219270.9, 2222105.32] |
| gumbel | $1,325,950$ | -$20,825,629$ | $41,651,261$ | $41,651,285$ | $\widehat{a} = 1.30 \times 10^6$ | [1135954.33, 1297516.1] | $\widehat{b} = 1.14 \times 10^6$ | [1134178.33, 1137906.37] |
| logis | $1,325,950$ | -$21,163,576$ | $42,327,155$ | $42,327,179$ | $\widehat{m} = 1.61 \times 10^6$ | [1016667.06, 1607560.29] | $\widehat{s} = 1.02 \times 10^6$ | [1015054.41, 1017887.2] |
| norm | $1,325,950$ | -$21,750,706$ | $43,501,417$ | $43,501,441$ | $\widehat{\mu} = 2.17 \times 10^6$ | [3220115.06, 2161041.21] | $\widehat{\sigma} = 3.22 \times 10^6$ | [3216296.98, 3223985.48] |

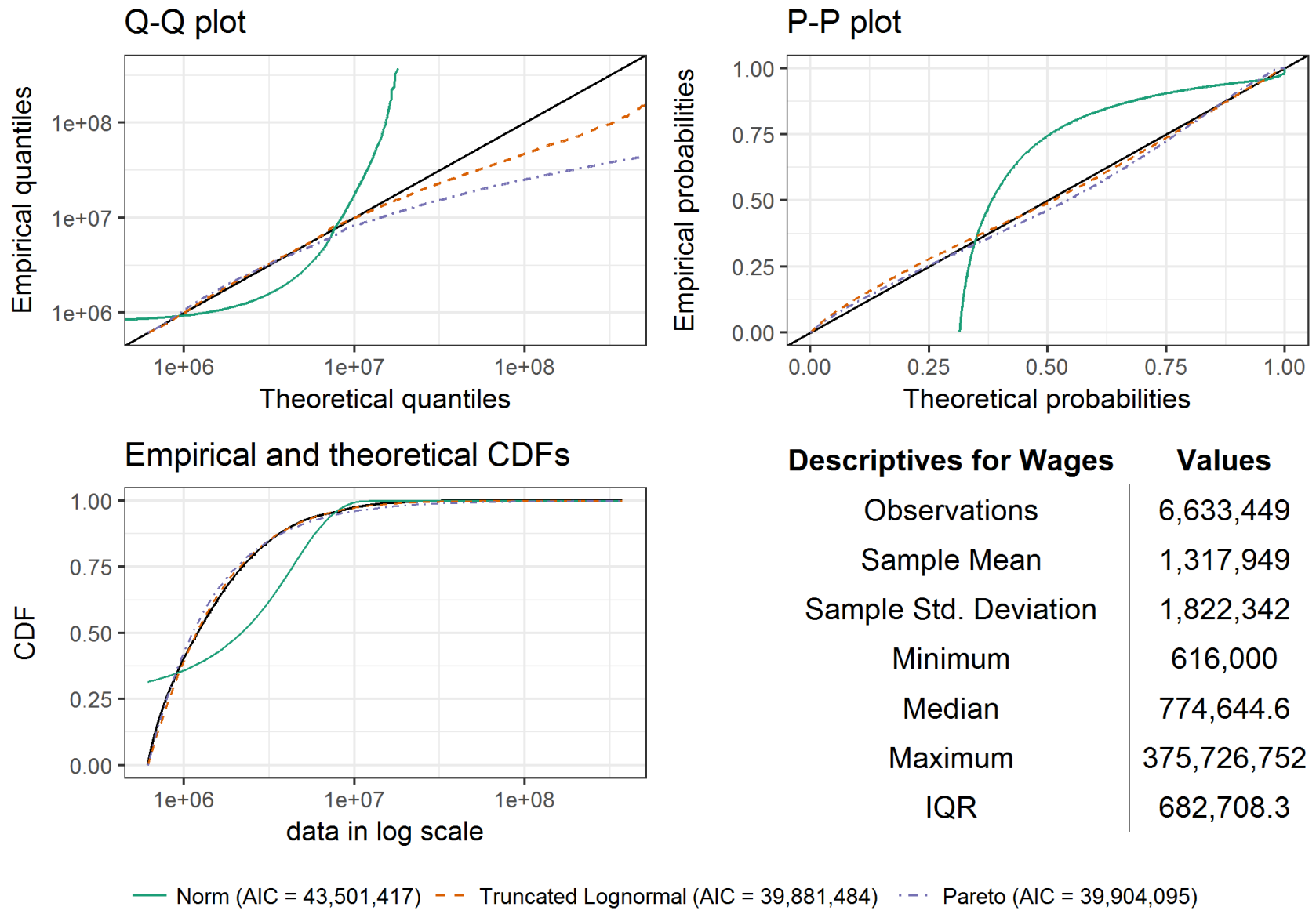

| Descriptives for Wages | Values |
|---|---|
| Observations | 6,633,449 |
| Sample Mean | 1,317,949 |
| Sample Std. Deviation | 1,822,342 |
| Minimum | 616,000 |
| Median | 774,644.6 |
| Maximum | 375,726,752 |
| IQR | 682,708.3 |

**Figure 7.** Diagnostic graphical comparison for the distributions of individual monthly wages (for workers living in municipalities with sizes above $n_{\min} = 287$), fitted by a truncated-lognormal, a Pareto, and a normal distributions, along with some descriptive statistics. Distributions that fit well the data should line up with the black solid line in the Q-Q and P-P plots. Clearly, the normal distribution (green line) is not a good fit for the distribution of monthly wages across workers. Ultimately, the relative best fit among many alternative distributions is given by the smallest AIC, according to which the (truncated) log-normal distribution is the preferred model for monthly wages among Colombian formal workers.

*Sizes* Figure 8 plots the full empirical complementary cumulative distribution function of municipality sizes. We have followed the methodology proposed by Clauset et al. (2009) to visualize and fit Pareto distributions. We observe in this empirical distribution a natural small-size scale, determined by the estimated minimum size, $\widehat{n_{\min}} \approx 287$ (vertical dashed line), above which the Pareto distribution is well fit (see Clauset et al. 2009 for how to estimate this parameter). We will carry out all our subsequent analyses on the municipalities above $\widehat{n_{\min}}$. Dropping the small-sized municipalities allows us to satisfy the assumption we used for Equation (12), that city sizes are Pareto distributed.[‡] Dropping municipalities that have less than 287 formal workers, means dropping from our analysis $80,526$ workers (only $1.2$ percent of total workers in our sample) and $564$ municipalities (approximately half of all municipalities).

[‡]Dropping the municipalities with the smallest sizes is typically done as this reduces the potential bias introduced by the fact that their formal employment is overrepresented by public servants whose wages are less determined by economic forces.

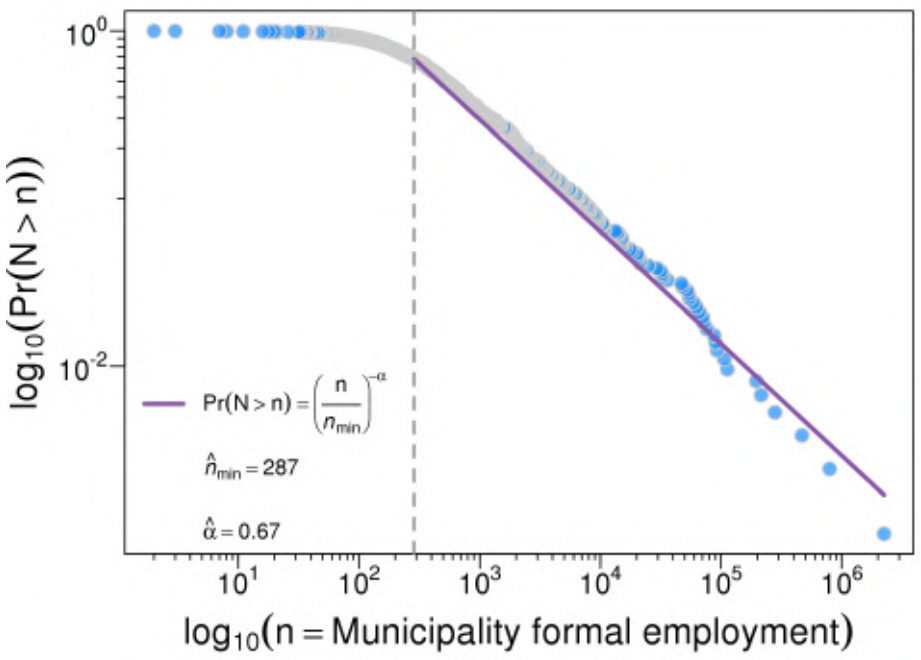

**Figure 8.** The complementary cumulative empirical distribution of number of workers across municipalities (blue circles) is well-fit by a Pareto distribution (solid purple line).

**Table 2.** Distributions fitted to municipality sizes. The list of the distributions are ordered from top to bottom by increasing AIC values.

| dist | numobs | loglik | AIC | BIC | Parameter 1 | C.I. | Parameter 2 | C.I. |
|---|---|---|---|---|---|---|---|---|
| powerlaw | 553 | -4, $733.51$ | $9,469.02$ | $9,473.33$ | $\widehat{\alpha} = 0.67$ | [0.61, 0.72] | | |
| trunclnorm | 553 | -4, $733.02$ | $9,470.05$ | $9,478.68$ | $\widehat{\ln(x_0)} = -22.96$ | [-50, 0.44] | $\widehat{\sigma} = 6.87$ | [3.46, 9.63] |
| truncweibull | 553 | -4, $733.08$ | $9,470.16$ | $9,478.79$ | $\widehat{a} = 0.04$ | [0.03, 0.11] | $\widehat{b} = 0$ | [0, 0] |
| trunccauchy | 553 | -4, $755.53$ | $9,515.06$ | $9,523.69$ | $\widehat{l} = 0.001$ | [0, 186.88] | $\widehat{s} = 428.74$ | [354.78, 531.55] |
| truncgamma | 553 | -4, $931.69$ | $9,867.38$ | $9,876.02$ | $\widehat{a} = 0$ | [0, 0] | $\widehat{\lambda} = 0.0000$ | [0, 0] |
| lnorm | 553 | -4, $942.60$ | $9,889.21$ | $9,897.84$ | $\widehat{\ln(x_0)} = 7.16$ | [7.04, 7.27] | $\widehat{\sigma} = 1.44$ | [1.35, 1.52] |
| weibull | 553 | -5, $115.31$ | $10,234.61$ | $10,243.24$ | $\widehat{a} = 0.50$ | [0.47, 0.55] | $\widehat{b} = 2,882.57$ | [2377.31, 3424.19] |
| gamma | 553 | -5, $299.62$ | $10,603.24$ | $10,611.87$ | $\widehat{a} = 0.31$ | [0.28, 0.34] | $\widehat{\lambda} = 0.0000$ | [0, 0] |
| truncgumbel | 553 | -5, $937.73$ | $11,879.45$ | $11,888.09$ | $\widehat{a} = 0.0002$ | [0, 3980.75] | $\widehat{b} = 10,684.55$ | [9403.29, 11307.02] |
| trunclogis | 553 | -6, $003.80$ | $12,011.61$ | $12,020.24$ | $\widehat{m} = 0.0001$ | [0, 5064.15] | $\widehat{s} = 10,458.00$ | [9169.54, 11040.73] |
| gumbel | 553 | -6, $188.90$ | $12,381.81$ | $12,390.44$ | $\widehat{a} = 2,211.40$ | [1374.19, 3351.65] | $\widehat{b} = 10,582.62$ | [9954.13, 11198.11] |
| norm | 553 | -7, $178.59$ | $14,361.17$ | $14,369.80$ | $\widehat{\mu} = 11,995.39$ | [4325.07, 22479.46] | $\widehat{\sigma} = 105,053.80$ | [100241.97, 111087.68] |

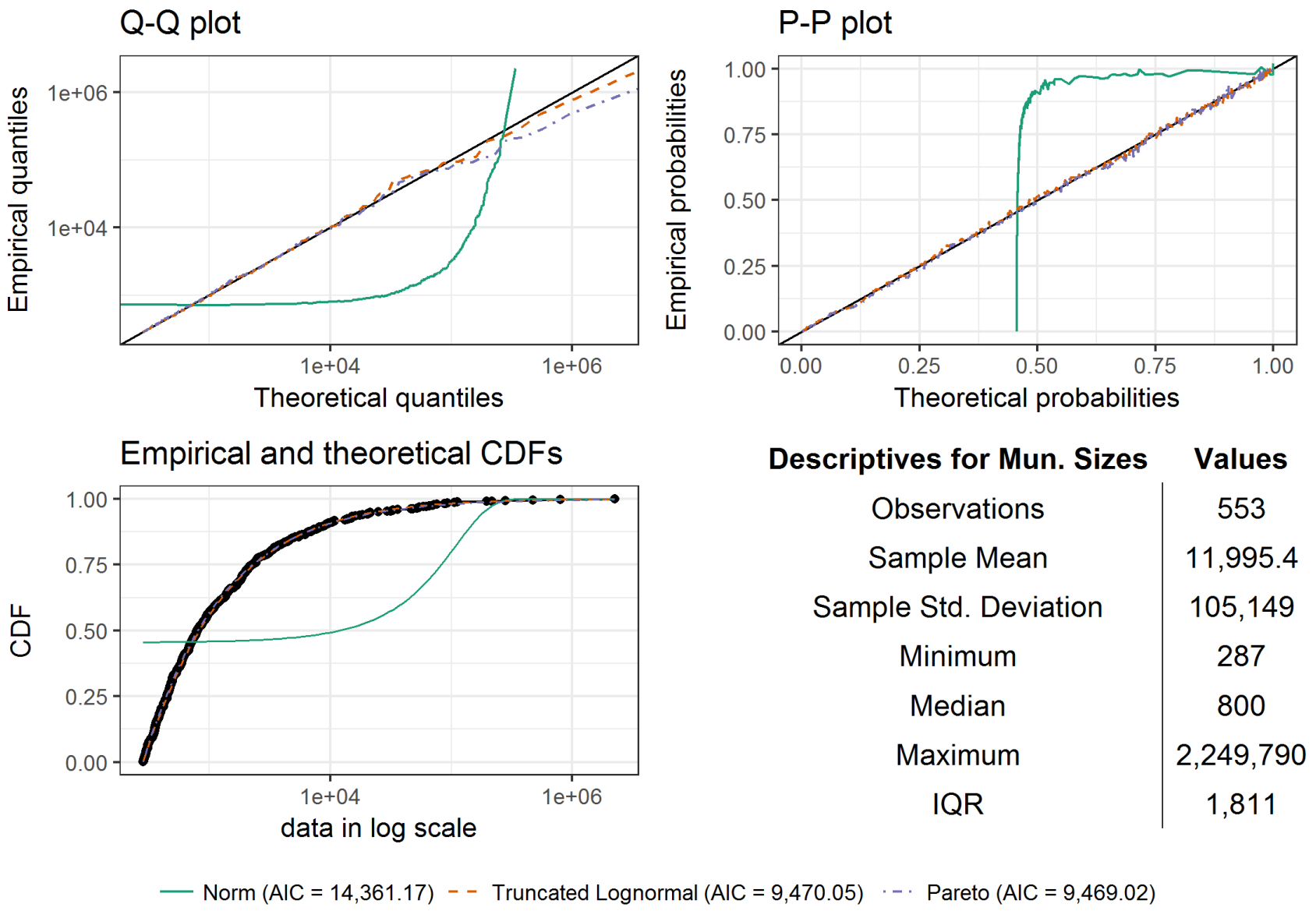

| **Descriptives for Mun. Sizes** | **Values** |
|---|---|
| Observations | 553 |
| Sample Mean | 11,995.4 |
| Sample Std. Deviation | 105,149 |
| Minimum | 287 |
| Median | 800 |
| Maximum | 2,249,790 |
| IQR | 1,811 |

**Figure 9.** Diagnostic graphical comparison for the distributions of municipality sizes (with sizes above $n_{\min} = 287$), fitted by a truncated-lognormal, a Pareto, and a normal distributions, along with some descriptive statistics. Distributions that fit well the data should line up with the black solid line in the Q-Q and P-P plots. Clearly, the normal distribution (green line) is not a good fit for the distribution of municipality sizes. Ultimately, the relative best fit among many alternative distributions is given by the smallest AIC, according to which the Pareto distribution is the preferred model for Colombian municipality sizes.